\documentclass{jfm}
\usepackage{amsmath,amsfonts,bm}
\usepackage[dvips]{graphicx}

\title{Statistics of Polymer Extension in a Random Flow with Mean Shear}

\author{M. Chertkov$^1$, I. Kolokolov$^{1,2}$, V. Lebedev$^{1,2}$, and
  K. Turitsyn$^{1,2}$}%

\affiliation{$^1$ Theoretical Division, LANL, Los Alamos, NM 87545, USA
\\[\affilskip]
$^2$Landau Institute for Theoretical Physics, Moscow, Kosygina 2,
  119334, Russia}

\date{\today}%{?? and in revised form ??}

\begin{document}

\maketitle

\begin{abstract}
Considering the dynamics of a polymer with finite extensibility placed in a chaotic flow
with large mean shear, we explain how the statistics of polymer extension changes with
Weissenberg number, ${\it Wi}$, defined as the product of the polymer relaxation time and
the Lyapunov exponent of the flow. Four regimes, of the ${\it Wi}$ number, are
identified. One below the coil-stretched transition and three above the coil-stretched
transition. Specific emphasis is given to explaining these regimes in terms of the
polymer dynamics.
\end{abstract}

%\pacs{83.80.Rs,47.27.Nz}

% LA-UR-04-6271

\maketitle

\underline{\bf Introduction.} Recently a number of experimental observations resolving the dynamics
of individual polymers (DNA molecules) in a permanent shear flow have been reported  by
\cite{99SBC}, see also \cite{01HSBSC}. These experimental results and the subsequent
theoretical/numerical study of \cite{00HSL} have focused on the analysis of the power spectral
density and simultaneous PDF of the polymer extension in the permanent shear, with fluctuations
driven by thermal noise. In another experimental development by Groisman \& Steinberg
(2000,2001,2004), a chaotic flow state called by the authors ``elastic turbulence'' was observed in
dilute polymer solutions. This flow consists of regular (shear-like) and chaotic components, the
latter being weaker. Resolving an individual polymer in this chaotic steady flow was the next
challenging but still accessible task achieved by \cite{04GCS}. The coil-stretch transition,
predicted by \cite{73Lum} (see also \cite{00BFL} and \cite{00Che}), was observed in direct
single-polymer measurements by \cite{04GCS}.

In this letter we present a theoretical analysis of the polymer extension statistics in a chaotic
flow with large mean shear, $s$, e.g. of the type corresponding to the elastic turbulence setup
described by Groisman \& Steinberg (2000,2001,2004). It is assumed that the flow is statistically
steady and thus the polymer attains a statistically steady distribution as well. We establish the
main features of the extension probability distribution, paying special attention to the PDF tails.

\begin{figure}%[tl]
   \includegraphics[width=1\textwidth]{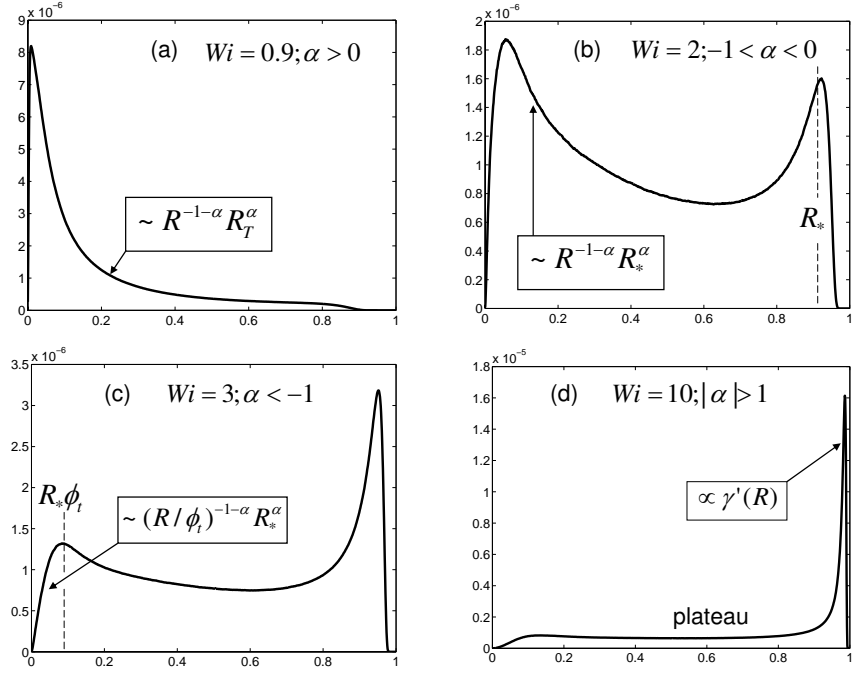}
 \caption{PDF of the polymer extension, $R$, measured in the units of
   maximal extension, for different values of the Weissenberg number, ${\it Wi}$,
   obtained from numerical simulations of the stochastic equations explained in the text.}
 \label{fig:PDF}
\end{figure}

The structure of this letter is as follows. We begin introducing
the basic dumb-bell-like equation governing dynamics of the
polymer end-to-end vector, $\bm R$, in a non-homogeneous flow.
Even though the prime interest of this letter is to describe the
statistics of the polymer extension, $R$ (the absolute value of
$\bm R$) the dynamics of $R$ is tightly linked to the angular
dynamics. This angular dynamics and the related statistics were
the subjects of our recent study, see \cite{04CKLTa}. A brief
explanation of these recent results, relevant to this letter
concludes our introduction. We then focus on the main subject - to
describe the statistics of the polymer extension. We formulate the
basic stochastic equation governing the dynamics of the polymer
extension. Then we analyze the structure of the extension PDF
which shows a strong dependence on the Weissenberg number, ${\it
Wi}$. We consider four cases corresponding to qualitatively
different PDF behaviors. We explain how the typical extension
depends on ${\it Wi}$ and examine the tails of the extension PDF,
for $R$ less than and larger than typical values. The structure of
the tails is complicated, consisting in some cases of a number of
asymptotic sub-intervals. We explain the dynamical origin of all
the sub-intervals. To illustrate our generic analytical results,
we present in Fig. \ref{fig:PDF} graphs, corresponding to the four
different regimes, obtained by direct numerical simulation made
within a model of short-correlated velocity statistics and of the
so-called FENE-P modeling of polymer elasticity.

\underline{\bf Model.} We consider a single polymer molecule which is advected by a
chaotic/turbulent flow (i.e. the polymer moves along a Lagrangian trajectory of the flow) and is
stretched by velocity inhomogeneity. The polymer stretching is characterized by the molecule's
end-to-end separation vector, ${\bm R}$, satisfying the following dumb-bell-like equation (see e.g.
\cite{77Hin,87BCAH}):
 \begin{equation}
 \partial_t R_i=R_j\nabla_j v_i-\gamma(R) R_i +\zeta_i\,.
 \label{model} \end{equation}
Here $\gamma$ is the polymer relaxation rate and $\zeta_i$ is the
Langevin force. The velocity gradient $\nabla_j v_i$ is taken at
the molecule position. The velocity difference between the polymer
end points is approximated in Eq. (\ref{model}) by the first term
of its Taylor expansion in the end-to-end vector. This
approximation is justified if the polymer size is less than the
velocity correlation length. The relaxation rate $\gamma$ in Eq.
(\ref{model}) is a function of the extension $R$ which varies from
zero upto a maximum value $R_\mathrm{max}$ corresponding to a
fully stretched polymer. We assume that the relaxation is Hookean
for $R\ll R_\mathrm{max}$, i.e. $\gamma(R)$ is well approximated
by a constant $\gamma_0$ there, while it diverges (the polymer
becomes stiff) for $R\to R_\mathrm{max}$.

 \begin{figure}%[tl]
  \includegraphics[width=0.6\textwidth]{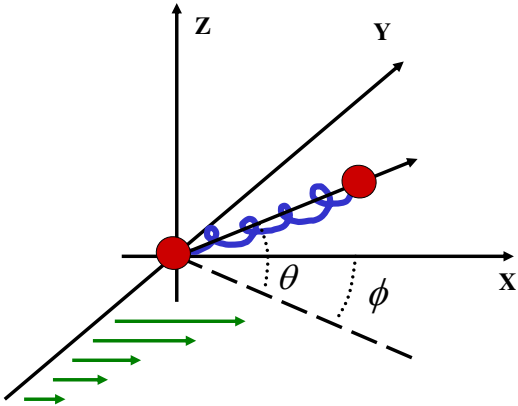}
 \caption{Scheme explaining polymer orientation geometry.}
 \label{fig:orient}
\end{figure}

We focus on the situation in which the effect of velocity fluctuations is stronger than that of
thermal fluctuations, so that the Langevin force $\bm\zeta$ in Eq. (\ref{model}) can be neglected.
We consider the case in which the steady shear flow is accompanied by weaker random velocity
fluctuations. This is also the setting realized in the elastic turbulence experiments by Groisman
\& Steinberg (2000,2001,2004). We choose the coordinate frame associated with the shear flow, as
shown in Fig. \ref{fig:orient}, where the mean shear flow is characterized by the velocity
$(sy,0,0)$ and $s$ is positive. Then the polymer end-to-end vector $\bm R$ is conveniently
parameterized by the spherical angles $\phi$ and $\theta$: $R_x=R\cos\theta\cos\phi$,
$R_y=R\cos\theta\sin\phi$, $R_z=R\sin\theta$. In terms of these variables, Eq. (\ref{model}) (with
the Langevin term omitted) transforms into the following set of equations:
 \begin{eqnarray} &&
 \partial_t{\phi} = - s\sin^2\phi + \xi_\phi \,,
 \label{phieq} \\ &&
 \partial_t{\theta} =- s
 \sin\phi\cos\phi\sin\theta\cos\theta + \xi_\theta \,,
 \label{thetaeq} \\ &&
 \partial_t \ln R=-\gamma(R)
 +s \cos^2\theta\cos\phi\sin\phi +\xi_\parallel \,,
 \label{Req} \end{eqnarray}
where $\xi_\phi$, $\xi_\theta$ and $\xi_\parallel$ are random variables related to the fluctuating
component of the velocity gradient. Note, that the angular (orientational) dynamics described by
Eqs. (\ref{phieq},\ref{thetaeq}) decouples from the dynamics of the polymer extension, $R$. Another
remark is that, at $\gamma=0$, Eq. (\ref{Req}) describes the divergence of neighboring Lagrangian
trajectories.

The character of the angular dynamics (closely related to the Lagrangian dynamics in the flow) was
discussed in detail by \cite{04CKLTa}. Typically, the polymer orientation fluctuates near its
preferred direction determined by the shear. Characteristic values of the fluctuations, both in
$\theta$ and $\phi$, are determined by the average value of the angle $\phi$, $\phi_t$, while the
average value of $\theta$ is zero. (We consider parametrization with the angles taken inside the
torus $-\pi/2<\phi,\theta<\pi/2$.) The value of $\phi_t$ is small due to assumed weakness of the
velocity gradient fluctuations $\xi$ in comparison with the shear rate $s$. Note that $\phi_t$ is
directly related to the value of the Lyapunov exponent of the flow, $\bar\lambda$: $\bar\lambda=
s\phi_t$. We make the natural assumption that the flow velocity is correlated at the time $\tau_t=
\bar{\lambda}^{-1}$. The random terms $\xi_\phi$ and $\xi_\theta$ in Eqs.
(\ref{phieq},\ref{thetaeq}) are relevant (i.e. comparable to their deterministic counter-parts)
only in the narrow angular region $|\phi|,|\theta| <\phi_t$. Outside this ``stochastic domain'' the
angular dynamics is mainly deterministic, i.e. the effect of the stochastic terms, $\xi_\phi$ and
$\xi_\theta$, is not essential there. The deterministic motion leads to polymer flipping (i.e.
reversing its orientation). The flips interrupt slower stochastic wandering near the
shear-preferred direction. The process of the deterministic/stochastic regimes alteration
(tumbling) is a-periodic.

\underline{\bf The Statistics of Polymer Extension.} We consider the case of a statistically steady
random flow, that leads to stationary statistics as a function of the angles, $\theta,\phi$, and of
the extension $R$. Our goal here is to describe the statistics of $R$ that emerges as a result of
the balance between elasticity-driven contraction and extension, caused by fluctuations in the
flow.

For the principal part of the dynamics, the basic dynamical equation Eq. (\ref{Req}) can
be simplified. First, the main contribution to the $R$ dynamics stems from the region of
small angles where $\sin\phi$ can be replaced by $\phi$ and $\cos\theta$ by unity.
Second, the term $\xi_\parallel$ is potentially important only in the stochastic region
where $\xi_\phi$ competes with $s\phi^2$. However, assuming $\xi_\phi\sim\xi_\parallel$,
we conclude that $\xi_\parallel$ is negligible in comparison with $s\phi$ there.
Therefore one arrives at:
 \begin{equation}
 \partial_t \ln R=-\gamma(R)+s \phi  \,.
 \label{ext} \end{equation}
Note that Eq. (\ref{ext}) is inapplicable when $R$ is close to its minimum value during a
flip (since the angles $\phi,\theta$ are of order unity there). Another remark is that
Eq. (\ref{ext}) is correct for $R\gg R_T$ ($R_T$ is the typical length of the polymer in
the absence of the flow) where the Langevin force can be neglected.

The statistics of $R$ is determined by the interplay of the two terms on the right-hand side of Eq.
(\ref{ext}). Since the average value of $s\phi$ is equal to the Lyapunov exponent, $\bar\lambda$,
the dimensionless parameter characterizing the statistics of the polymer extension is the
Weissenberg number, ${\it Wi}\equiv\bar\lambda/\gamma_0$, which grows with the strength of the
shear, or/and, of the velocity fluctuations. At ${\it Wi}=1$, when the two terms on the right hand
side of Eq. (\ref{ext}) balance each other, the system undergoes the so-called coil-stretched
transition, see \cite{69Lum,73Lum,00BFL,00Che} and \cite{01BFL} for details. We find, however, that
in the specific case of strong shear considered in the letter additional qualitative changes in the
PDF of $R$ occur at ${\it Wi}>1$ so that the overall picture is richer than the case of isotropic
velocity statistics. Below we describe characteristic features of the extension PDF as a function
of ${\it Wi}$.

To illustrate our generic analytical results we plot in Fig. \ref{fig:PDF} four graphs of
the extension PDF obtained by numerical simulations based on modification of Eq.
(\ref{ext}), $\partial_t \ln R=-\gamma+s \sin\phi\cos\phi$, which allows the correct
reproduction of the flips, and Eq. (\ref{phieq}) with the stochastic term $\xi_\phi$
chosen to be $\delta$-correlated in time. The simulations were done with
$\gamma(R)=\gamma_0/(1-R^2/R^2_\mathrm{max})$, corresponding to the so-called FENE-P
model of the polymer elasticity, see e.g. \cite{87BCAH}.

\underline{${\it Wi}<1$, $\alpha>0$.} We begin by discussing the case ${\it Wi}<1$, for
which the polymer is only weakly stretched. Then, typically, the molecule stays in the
``coil'' state characterized by the thermal size, $R_T$, which emerges as the result of a
balance between the Langevin driven extension and contraction (relaxation) related to
polymer elasticity. Let us recall that due to a large number of monomers in the polymer
molecule, the thermal noise induced length, $R_T$, is much smaller than the maximal
polymer extension, $R_\mathrm{max}$.

At the scales larger than $R_T$ the thermal noise is irrelevant and the extension
dynamics is described by Eq. (\ref{ext}). Being interested in the statistics for large,
$R\gg R_T$, for deviations from the typical value, $R_T$, one comes to a problem that was
already analyzed in detail by \cite{00BFL,00Che,01BFL}, in which it was shown that the
extension PDF has an algebraic tail:
 \begin{equation}
 P(R)\propto R^{-1-\alpha}\,,
 \label{alg} \end{equation}
with $\alpha>0$. Eq. (\ref{alg}) holds at $R_\mathrm{max}\gg R\gg R_T$ where $\gamma(R)$
weakly deviates from $\gamma_0$. The situation is reflected in Fig. \ref{fig:PDF}a, where
the algebraic tail is clearly seen. The positive value of the exponent $\alpha$ in Eq.
(\ref{alg}) guarantees that the normalization integral $\int dR\, P(R)$ converges in the
region $R\gg R_T$. Thus, the normalization coefficient in Eq. (\ref{alg}) is $\sim
R_T^{-\alpha}$. The exponent $\alpha$ decreases as the Weissenberg number, ${\it Wi}$,
increases, and it crosses zero at the coil-stretch transition, where ${\it Wi}=1$.

The algebraic tail (\ref{alg}) is related to a long (compared to the correlation time $\tau_t$)
process of polymer extension from the typical value $R_T$ to the current value of the extension,
$R\gg R_T$. Note that this long extension does not mean that (during the process of extension) the
right hand side of Eq. (\ref{ext}) is always positive since $\phi$ fluctuates and $s\phi$ is larger
than $\gamma$ only on average. Moreover, the process consists of alternating stochastic and
deterministic portions (polymer flips during the later ones). In spite of the fact that $R$
decreases during the first half of the flip, the initial extension is restored (returns to its
initial value) during the second half of the flip. Overall, the flips do not influence the
extension process. The probability $W$ of this a-typically long extension process depends
exponentially on its duration $T$, since $W$ is a product of independent probabilities, each
representing a sub-processes of duration $\tau_t$. This gives the following estimate: $\ln W\sim
-T/\tau_t$. On the other hand, in accordance with Eq. (\ref{ext}), $\ln(R/R_T)\sim T/\tau_t$.
Combining the two estimates, we arrive at the algebraic tail (\ref{alg}) for the PDF of extension,
$P=dW/dR$.

\underline{${\it Wi}>1$, $-1<\alpha<0$.} Above the coil-stretch transition, when
$\bar\lambda$ exceeds $\gamma_0$, the polymers become strongly extended. In this
stretched state the typical size of the polymer, $R_*$, is much larger than $R_T$.
Considering the stationary average of Eq. (\ref{ext}) one finds that
$\gamma(R_*)=\bar\lambda$.

The left tail of the PDF, corresponding to $R_T\ll R\ll R_*$, is governed by the same
algebraic law (\ref{alg}). However, now $\alpha<0$, meaning that the normalization
integral $\int dR\, P(R)$ gets a major contribution for $R\sim R_*$. Therefore, restoring
normalization, one derives $P\sim R_*^\alpha R^{-1-\alpha}$ for the $R\ll R_*$ tail. The
transition from positive to negative $\alpha$ corresponds to important change in the
nature of the dynamical configuration corresponding to the algebraic tail: extension, as
a typical process for $\alpha>0$, is replaced by contraction for negative $\alpha$, so
that initially typical extension, $\sim R_*$, contracts through a long, $T\gg\tau_t$,
(multi-tumbling) evolution. (Physical arguments, clarifying the origin of the algebraic
tail, are identical to the ones presented above for ${\it Wi}<1$.)

The right tail, corresponding to extreme extensions, $R_\mathrm{max} -R \ll R-R_*$, can
also be explained in some general terms. The domain of extreme deviation is characterized
by extremely fast relaxation, so that the term on the left hand side of Eq. (\ref{ext})
can be neglected. As a result, $R$ and $\phi$ are related to each other locally:
$\gamma(R)= s\phi$. Moreover, one finds that, because of the fast nature of the polymer
relaxation in the extreme case, $\phi$ simply follows the respective random term in Eq.
(\ref{phieq}), i.e. the term on the left hand side of Eq. (\ref{phieq}) can also be
neglected resulting in, $s\phi^2=\xi_\phi$. In other words, the extreme configuration is
produced through fast anti-clockwise revolution of the polymer to a large (in comparison
with the typical value $\phi_t$) positive angle, $1\gg\phi\gg\phi_t$, driven by anomalous
fluctuations in $\xi_\phi$. Recalculating the PDF of $\xi_\phi$ into $P(R)$ one arrives
at
 \begin{equation}
 P(R)=2 s^{-1}\gamma\gamma' P_\xi(\gamma^2/s) \,,
 \label{prep1} \end{equation}
where $P_\xi$ is the simultaneous PDF of the velocity gradient term $\xi_\phi$. Note that
the asymptotic expression given by Eq. (\ref{prep1}) is not restricted to the case
considered in this subsection but applies generically to the description of extreme
polymer extensions in all regimes.

\underline{${\it Wi}>1$, $\alpha<-1$.} Once $\alpha$ crosses $-1$, $R_*$ becomes maximum
of the extension PDF, $P(R)$. This modification in the PDF shape is accompanied by the
emergence of a plateau on the left from the maximum (see Fig. \ref{fig:PDF}c), associated
with an additional contribution to the PDF related to the deterministic angular dynamics.

Let us explain the origin of the plateau. For angles which are smaller than unity but
larger than $\phi_t$, $R$ and $\phi$ satisfy, $\partial_t\ln R=s\phi$ and
$\partial_t\phi=-s\phi^2$, as follows from Eqs. (\ref{phieq},\ref{ext}). Integrating
these equations one arrives at, $R=A|t-t_0|$ where $t_0$ and $A$ are constants, the
latter being estimated by $A\sim R_*/\tau_t$. Taking into account the fact that the time
$t_0$ is homogeneously distributed (due to the assumed homogeneity of the velocity
statistics), and recalculating the measure $dt_0$ into the PDF of $R$, one arrives at
$P(R)=C/R_*$ ($C$ is an $R$-independent constant of order unity) corresponding to the
plateau seen in Fig. \ref{fig:PDF}c.

The ``deterministic'' contribution to $P(R)$, $\sim 1/R_*$, discussed above, does not
cancel the ``stochastic'' one $\sim R_*^{\alpha} R^{-1-\alpha}$ corresponding to the long
contraction which starts at $R_*$. In fact these contributions co-exist. One finds that
for $-1<\alpha<0$, the stochastic contribution dominates, in full agreement with the
above discussion of Fig. \ref{fig:PDF}b. When $\alpha$ becomes smaller than $-1$, the
situation is reversed and the deterministic contribution dominates.

The plateau extends from $R\sim R_*$ down to $R\sim R_*\phi_t$, where $R_*\phi_t$ is the
smallest value of $R$ one can achieve under the condition that when the flip begins the
initial extension is $R_*$. Note however, that if the initial extension is smaller than
$R_*$, a deterministic flip could bring it to values that are smaller than $R_*\phi_t$.
Therefore, to understand even smaller values of $R$, $R< R_*\phi_t$, one should consider
flips which begin with an anomalously small initial value $R_0$, $R_0<R_*$ (prepared by
some preliminary and long stochastic processes, of the type discussed above). The
probability density of achieving $R_0$ during the preparatory stage is estimated
according to Eq. (\ref{alg}): $\sim R_*^{\alpha} R_0^{-1-\alpha}$. On the other hand, any
$R_0$ that lies between $R$ and $R/\phi_t$ transforms dynamically as a result of a fast
flip into the current value of extension $R$ with the same $R$-independent probability,
which now can be estimated as $\sim 1/R_0$. Therefore, one arrives at the following
estimate for the PDF of $R$ at $R<R_*\phi_t$:
 \begin{equation}
 P\sim\int_{R}^{R/\phi_t}\frac{dR_0}{R_0}\times \frac{R_*^{\alpha}}
 {R_0^{1+\alpha}}\sim
 R_*^{-|\alpha|}\left(\frac{R}{\phi_t}\right)^{|\alpha|-1}.
 \label{extra} \end{equation}
Eq. (\ref{extra}) explains the probability decrease at the smallest $R$ seen in Fig.
\ref{fig:PDF}c.

Let us observe the bump in Fig. \ref{fig:PDF}c separating the region of the plateau and
the smallest $R$ region of the probability decay. To explain the bump, one simply needs
to account for the angular nonlinearity (with respect to $\phi$ and $\theta$) in the
estimate of the plateau just discussed.

\underline{${\it Wi}\gg 1$, $\alpha\ll-1$.} The larger ${\it Wi}$ is, the closer $R_*$
approaches $R_\mathrm{max}$. Then the condition of fast relaxation,
$R\gamma'(R)\gg\gamma$, which has already allowed us to analyze the extreme asymptotics
(\ref{prep1}), also applies to the region in the vicinity of $R_*$. The smallness of the
ratio $\gamma/(R\gamma')$ suggests that the left hand side of Eq. (\ref{ext}) can be
replaced by zero. One arrives at $\gamma(R)=s\phi$, which makes it easy to express the
PDF of $R$ through the simultaneous PDF of $\phi$:
 \begin{equation}
 P(R)\sim s^{-1} \gamma'(R)P_\phi(\gamma/s) \,,
 \label{prep} \end{equation}
where it is also assumed that $\gamma(R)/s< \phi_t$. In this special domain of $R$ and
$\phi$, variations of $P_\phi$ are slow, so that the main dependence on $R$ in Eq.
(\ref{prep}) is due to the factor $\gamma'(R)$. Eq. (\ref{prep}) applies to the left (for
smaller values of $R$) of $R_*$ provided the parameter $R\gamma'(R)/\gamma$ is large. In
this domain, the PDF can be estimated as $P\sim \gamma'(R)/\gamma(R_*)$. At even smaller
values of $R$, the PDF has a plateau, $P\sim \gamma_0/[R_*\gamma(R_*)]$ (which
generalizes our previous result to the ${\it Wi}\gg 1$ case). This explains the complex
behavior of the PDF of $R$ shown in Fig. \ref{fig:PDF}d.

\underline{\bf Conclusions.} In this second work devoted to the statistics of polymer
molecules in a chaotic flow with mean shear, we focused on analyzing the statistics of
the polymer extension $R$. The PDF of $R$ demonstrates complex and rich ${\it
Wi}$-dependent behavior related to a number of distinct processes governing polymer
dynamics. We observed that the typical value of the extension associated with the
stochastic wandering of the polymer orientation around the special shear-preferred
direction increases with ${\it Wi}$. In the region of maximal stretching, near
$R_\mathrm{max}$, the major contribution to the PDF originates from processes
characterized by fast adjustment of the polymer extension to the current value of the
polymer's angular degree of freedom. We also identified special contributions to the PDF
tails associated with fast (deterministic) flips and long (stochastic) extension or
contraction processes. Encouraged by qualitative agreement of our results with the newest
experimental data of \cite{04GCS}, we anticipate that the rich zoo of theoretical
predictions presented in the paper will be helpful for guiding and testing future
experimental work in this field.

We thank A. Celani and V. Steinberg for stimulating discussions and G. D. Doolen for
useful comments. Support of RSSF through personal grant (IK), RFBR grant 04-02-16520a
(IK,VL and KT), and a grant from the Dynasty Foundation (KT) are acknowledged.

\end{document}